\documentclass[12pt,aps,pre,amsmath,amsfonts,amssymb,floatfix,longbibliography,superscriptaddress,notitlepage,onecolumn]{revtex4-1}

\usepackage{amsfonts}
\usepackage{amsmath}
\usepackage{dcolumn}
\usepackage{epsfig}
\usepackage{graphicx}% Include figure files
\usepackage{subfigure}
\usepackage{dcolumn}% Align table columns on decimal point
\usepackage{bm}% bold math
\usepackage{times}
\usepackage{xcolor}
\usepackage{float}
\begin{document}

%\bibliographystyle{science}
%\preprint{APS/123-QED}
\title{Non-trivial Resource Amount Requirement in the Early Stage for Containing  Fatal Diseases}
%\title{Abrupt Phase Transitions Induced by Insufficient Resource in Containment of Fatal Diseases}
%\title{Critical Amount of Resource for Containing  Fatal Diseases in Real Contact Networks}
\author{Xiaolong Chen}\thanks{These three authors contributed equally}
\affiliation{School of Data and Computer Science, Sun Yat-sen University, Guangzhou 510006, China}
\affiliation{Web Sciences Center, University of Electronic Science and Technology of
China, Chengdu 611731, China}
\affiliation{Big Data Research Center, University of Electronic Science and Technology of China, Chengdu 611731, China}
\author{Tianshou Zhou}\thanks{These three authors contributed equally}
\affiliation{School of Mathematics, Sun Yat-sen University, Guangzhou 510006, China}
\author{Ling Feng}\thanks{These three authors contributed equally}
\affiliation{Institute of High Performance Computing, A*STAR, 138632 Singapore}
\author{Junhao Liang}
\affiliation{School of Mathematics, Sun Yat-sen University, Guangzhou 510006, China}
\author{Fredrik Liljeros}
\affiliation{Department of Sociology, Stockholm University, Stockholm, Sweden}
\author{Shlomo Havlin}
\affiliation{Minerva Center and Department of Physics, Bar-Ilan University, Ramat Gan, Israel}
\author{Yanqing Hu}\email{huyanq@mail.sysu.edu.cn}
\affiliation{School of Data and Computer Science, Sun Yat-sen University, Guangzhou 510006, China}
\affiliation{Big Data Research Center, University of Electronic Science and Technology of China, Chengdu 611731, China}

\date{\today}

\begin{abstract}

During an epidemic control, the containment of the disease is usually achieved through increasing devoted resource to shorten the duration of infectiousness. However, the impact of this resource expenditure has not been studied quantitatively. Using the well-documented cholera data, we observe empirically that the recovery rate which is related to the duration of infectiousness has a strong positive correlation with the average resource devoted to the infected individuals. By incorporating this relation we build a novel model and find that insufficient resource leads to an abrupt increase in the infected population size, which is in marked contrast with the continuous phase transitions believed previously. Counterintuitively, this abrupt phase transition is more pronounced in the less contagious diseases, which usually correspond to the most fatal ones. Furthermore, we find that even for a single infection source, public resource needs to meet a significant amount, which is proportional to the whole population size to ensure epidemic containment. Our findings provide a theoretical foundation for efficient epidemic containment strategies in the early stage.

\end{abstract}

\maketitle

In recent 20 years, epidemic outbreaks such as Ebola, SARS, H1N1, HIV etc., had detrimental impacts globally on both public health and social activities \cite{AidsEconomicCost,gubler2002Epidemic,sachs2002economic,stewart2004cost}. Understanding of the mechanism of epidemic spreading involved with human activities is a crucial issue\cite{liljeros2001web,helbing2013globally,hufnagel2004forecast,Vespignani2006PNASTransportation,liljeros2003sexual}.
Previous  studies about epidemic spread have mostly focused on the impact of contact network structure  \cite{pastor2001epidemic,parshani2010epidemic,lagorio2011quarantine,boguna2003absence,brockmann2013hidden}, individual mobility \cite{hufnagel2004forecast,Vespignani2006PNASTransportation} or modelling of the spreading process by taking account of many complex aspects\cite{wohlfeiler2005using,ferguson2006strategies,Holme2011Plossimulated,pandey2014strategies,SEIR}. Usually, the spreading power has been considered as constant though the whole spreading process. However disease needs human subjects to spread, the spreading power is not constant as commonly assumed, but highly influenced by the resources invested to fight the spreading. The effect of resource-dependent spreading power has been largely overlooked in the past, in contrary to the real dynamics.

The devoted resources can effectively shorten the duration of infectiousness and consequently decrease the spreading power. The duration of infectiousness not necessarily is the same as duration of being sick. Infected individuals may become infectious before the show any symptoms such as the yearly flue. Some infections may not show any symptoms at all, such as Chlamydia Trichomatis \cite{heymanncontrol}. The possibility of shortening the time of contagiousness varies for different diseases and can be medical, such as quickly identifying, quarantining and curing the infected population, as well as social, such as a good insurance for loss of income which may motivate the the staff at a workplace to stay at home when sick and hence shorten infectiousness at the workplace. Common to most methods is that  their implementation have a cost. Hence, it highly depends on the devoted resource. This is in particular critical in the case of fatal diseases, which usually attract more public attention and more resources in their containment.

\section{Results}

The duration of infectiousness is the key point for the disease spreading power. In order to illustrate this effect, here, we define the recovery rate $\mu$ as the probability of that the infected individuals are out of the duration of infectiousness. Regarding the relationship between the devoted resource $R$ and the recovery rate $\mu$, on one hand, if the amount of resource is limited, i.e., $R$ is a fixed value, the overall recovery rate of the system is closely related to the infected population size $\rho$. Namely, the larger the infected population the less resource per infected individual would share, and hence a smaller chance for the infected individual to recover. On the other hand, if the quantity of the infected populations is constant at the moment, the recovery rate would be improved by properly increasing the investment of resource. It implies that the recovery rate function $\mu(R,\rho)$ is a monotonic increasing function of $R$ and decreasing function of $\rho$, such that more total resource or less infected people sharing the resource results in faster recovery rate. More over, the value of $\mu$ should be constrained between 0 and 1 when $R$ and $\rho$ is positive. These two basic properties reflect the relations between $R$, $\rho$ and recovery rate.

Motivated by our empirical finding from cholera (See section I and II in SI), we consider the recovery rate as the following:
\begin{equation}
\mu(R,\rho) = e^{- c\rho/R},
\label{EN:mu}
\end{equation}
where $c$ is the coefficient representing the relative importance of $R$ and $\rho$.
Without losing any generality, {\it we further prove that for any function whose value is in $(0,1)$ and is a monotonic increasing function of $\frac{R}{\rho}$ (average shared resource per infectious), the main results in the paper will keep the same in both homogeneous and heterogeneous networks} (See SI section III). Thus, without loosing the generality, in this paper, we only discuss Eq. (\ref{EN:mu}). Introducing the effect of resource $R$ into the susceptible-infected-susceptible dynamics\cite {pastor2001epidemic,parshani2010epidemic,lagorio2011quarantine,boguna2003absence}, we build a novel spreading model, based on the time varying recovery rate dependent on the average resource that each infected individual gets. Formally, in a system of $N$ individuals embedded in a contact network, an infected individual (or node) has a probability $\beta$ to spread the disease to a neighbour to which it shares a link. At the same time, the infected node has a probability of $\mu(t)=e^{-\rho(t)/R}$ (corresponds to Eq. (\ref{EN:mu}) with $c=1$) to recover to a susceptible state, such that it does not carry the disease until it is reinfected again. Therefore the probability $p_i(t)$ that a node $i$ is infected at time $t$ can be  described by the following dynamical equation
\begin{equation}
\frac{dp_i(t)}{dt}=(1-q_i(t))(1-p_i(t)) - e^{-\rho(t)/R} p_i(t),
\label{EN:infection}
\end{equation}
where $q_i(t)=\prod_{j=1}^N(1-a_{ij}\beta p_j(t))$ is the probability that node $i$ is not infected by any of its neighbours. The parameter $a_{ij}$ denotes the adjacency matrix of the contact network and has a binary value. It is 1 when node $j$ shares a link with node $i$, and 0 otherwise. The infected population fraction is $\rho(t)=\frac{1}{N}\sum_{i=1}^N p_i(t)$ which is the infected fraction of the total population in time $t$. One of the main quantities we wish to address using this equation is to study the infected population size in the steady state, i.e. $\rho(\infty)$, and how it changes with the total amount of resource $R$.

For further analysis of Eq. (\ref{EN:infection}),we assume that the underlying network has degree distribution $P(k)$. With the mean field approximation, the dynamical equation Eq. (\ref{EN:infection}) can be reduced into
\begin{equation}
\frac{d\rho(k,t)}{dt}=k \beta \Theta (1-\rho(k,t)) - e^{-\rho(t)/R} \rho(k,t)
\label{Heterodynamics}
\end{equation}
Here, $\rho(k,t)$ is the infected fraction of nodes with degree $k$ . $\Theta$ denotes the probability that a randomly chosen edge links to an infected node. When the degree correlation of the network is not taken into consideration, $\Theta$ takes the form
\begin{equation}
\Theta = \sum_k \frac{k P(k) \rho(k,t)}{\langle k \rangle}
\end{equation}
Now the total infected fraction $\rho(t)$ can be given by $\rho(t) = \sum_k P(k) \rho(k,t)$. In general, $\rho(\infty)$ cannot be analytically deduced from Eq.(\ref{Heterodynamics}). However, in the following two typical cases, $\rho(\infty)$ can be given analytically. The first case is that each node has almost the same degree $\langle k \rangle$, thus the degree distribution $P(k)\approx \delta (k-\langle k \rangle)$  , for which we say the network is homogeneous. The second case is that the degree distribution takes the form of a power-law: $P(k)= \frac{2 m^2}{k^3}$ (we mainly consider the case of exponent 3 for simplicity), where $m$ is the minimum degree of a node, for which we say the network is heterogeneous. Denote $\varphi[\rho(t)] = \frac{2 e^{-\rho(t)/R}}{\beta \langle k \rangle}$, we can get the self-consistent equation of $\rho(\infty)$ as:

\begin{equation}
\rho(\infty)=1-\frac{1}{2}\varphi[\rho(\infty)] ,
\label{fixeq1}
\end{equation}
for homogeneous networks, and
\begin{equation}
\rho(\infty)=\frac{2\{e^{\varphi[\rho(\infty)]}-\varphi[\rho(\infty)]-1\}}{\{e^{\varphi[\rho(\infty)]}-1\}^2} ,
\label{fixeq2}
\end{equation}
for heterogeneous networks with exponent 3. The final infected fraction $\rho(\infty)$ is given by the fixed point of the above Eq. (\ref{fixeq1}) or Eq. (\ref{fixeq2}) (See SI section III for details).

%\subsection{Critical Resource Amount}

\begin{figure}[H]
 \center\includegraphics[width=1\linewidth]{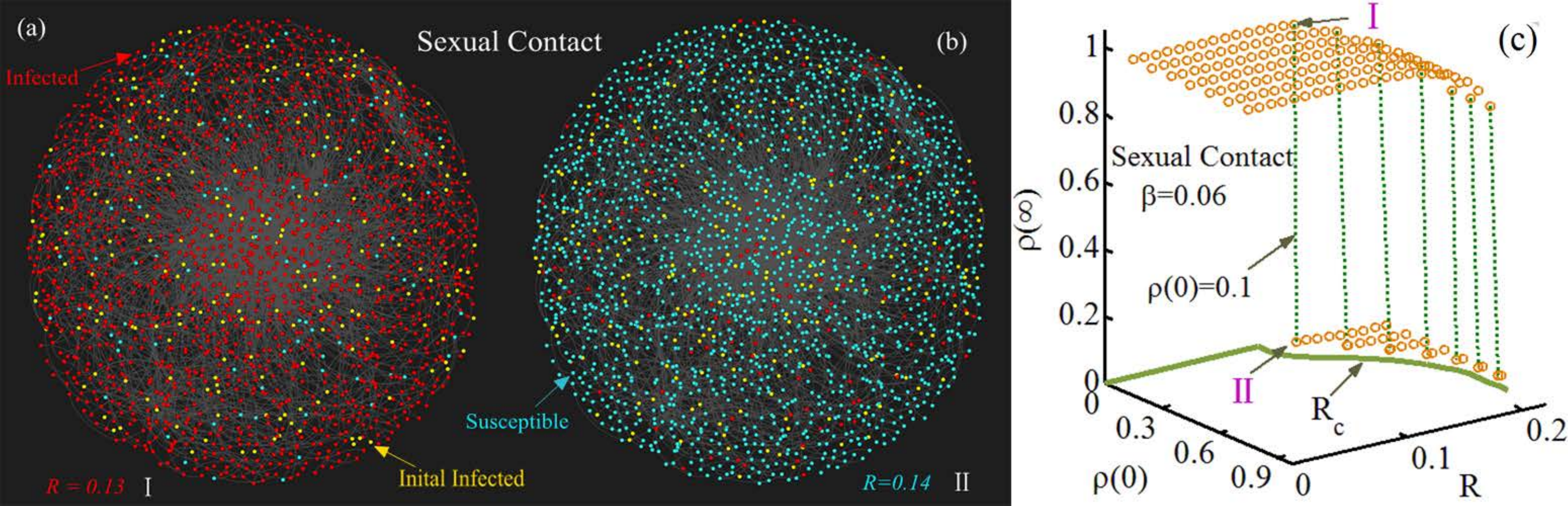}
  \caption{Catastrophic epidemic spreading due to inadequate resource $R$ simulated on the real sexual contact network. a and b, The simulation is carried out  with $\rho(0)=0.1$ and $\beta=0.06$. The yellow, red and blue nodes are the initially infected, finally infected and susceptible nodes respectively. Near the critical point $R_c\approx 0.135$, with a small change of $R$ from 0.13 to 0.14, the infected population decreases drastically. c. The critical amount of resource $R_c$ that separates the two phases is dependent on the initial fraction of infected population $\rho(0)$. There is a critical $R_c$ such that when $R<R_c$ the disease would spread to a significant fraction of the population as shown in region I; whereas when $R>R_c$ the disease would be well contained within a negligible fraction of population as shown in the controllable region II. As  $\rho(0)$ increases, the minimum amount of resource $R_c$ also increases to ensure the disease is not widespread. The open circles and dotted lines are simulation and theory results respectively.}
\label{FIG:Sextual}
\end{figure}

\begin{figure}[H]
\centering\includegraphics[width=1\linewidth]{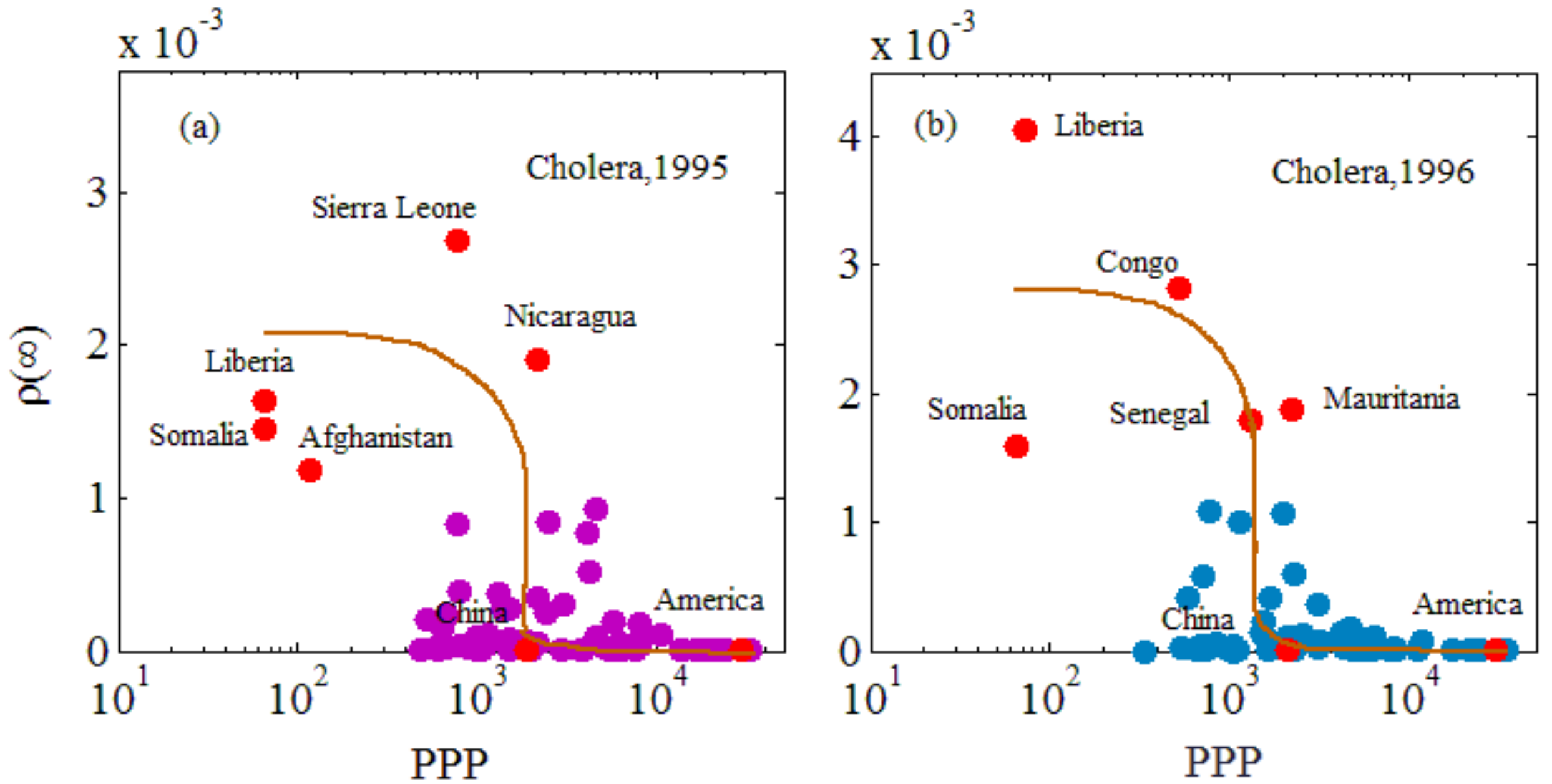}
\caption{Cholera infected density $\rho(\infty)$ vs. PPP in 1995 (a) and 1996 (b). Each point corresponds to a particular country. We can see the abrupt change of $\rho(\infty)$ around PPP=$10^{3}$ dollars, which correspond to our finding of first order phase transition. The curves indicate the approximate decaying trend in the data points.}
\label{FIG:cholerafirstorder}
\end{figure}

To investigate the impact of resources, we simulate the model, Eq. (\ref{EN:infection}), on the real sexual contact network \cite{liljeros2001web,Holme2011Plossimulated}.
%New Orleans' Facebook network in 2009 \cite{Viswanath2009Facebook}. This is the online social network in the early years of Facebook in the city of New Orleans, therefore could be a approximation of real contact network.
Surprisingly, the model shows a bimodal outcome depending on the resource $R$ available. As shown in Fig.~\ref{FIG:Sextual}, the final infected population is either extremely widespread to a considerable fraction of the network, or only a few nodes are being infected.
At a critical point of resource amount $R_c$, a tiny change in the resource variable $R$ would make a huge difference in $\rho(\infty)$. Such abrupt transition indicates that adequate critical resource is needed in fighting the disease spreading, since it could bring a potentially catastrophic epidemic down to a tiny outbreak (see Fig.~\ref{FIG:Sextual}). This is a signature of first order phase transition related to the resource available. It implies that the public resource expenditure has a critical behavior: When it is above this critical value, the disease could be effectively eradicated or contained, otherwise it cannot contain the outbreak but only slightly reduces the infected population size. {\it This abrupt transition is our first main finding}. In addition to this example of abrupt transition found in real sexual contact network, we also find the same abrupt transition due to resource $R$ in several other typical social networks, including New Orleans' Facebook network, Twitter network and Weibo network, which could approximate real contact networks, as well as artificial networks like Erd\"os-R\'enyi and Scale Free networks (see section I and Fig.S7 in SI). Most importantly, this kind of abrupt transition can be detected in the real data of cholera see Fig.~\ref{FIG:cholerafirstorder}.

\begin{figure}[H]
 \centering\includegraphics[width=1\linewidth]{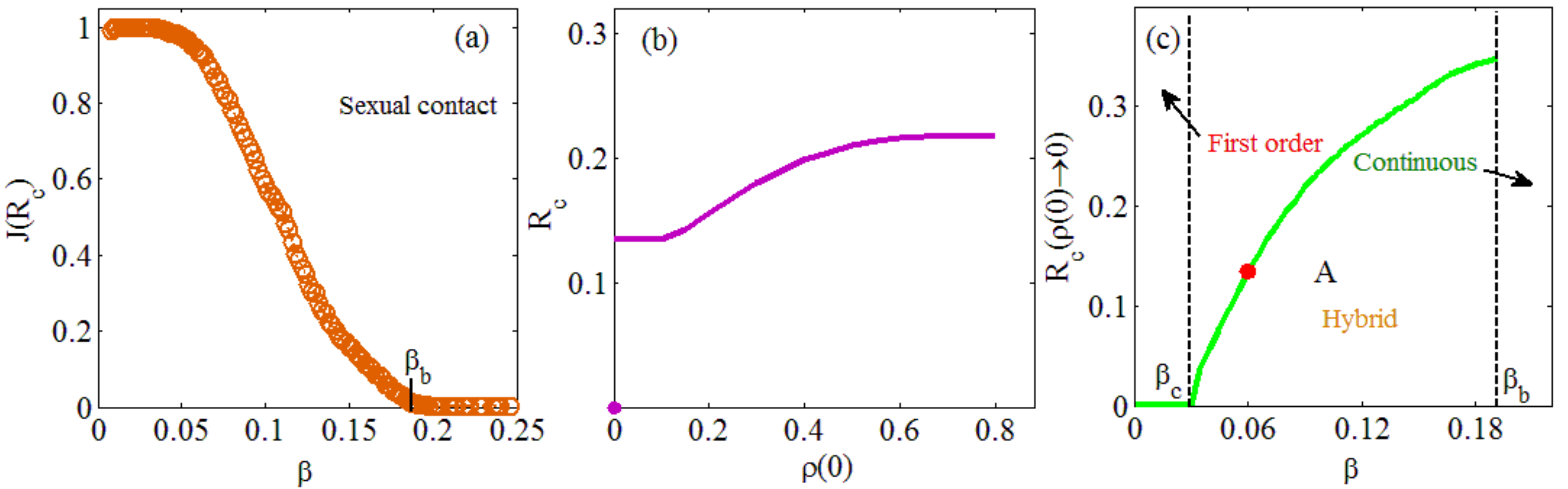}
 \caption{Non-trivial critical resource amount and multi-phase behaviors in real sexual contact network. (a) Shows the size of jump $J(R_c)$ in $\rho(\infty)$ at the critical resource value $R_c$ as a function of beta. When the infection rate $\beta$ is larger than  $\beta_b$, the catastrophic jump behavior disappears, switching to a continuous phase. (b)
$R_c$ is significantly larger than 0, even with close to 0 fraction of initial spreaders. Here $\beta=0.06$. (c) Shows the critical resource $R_c$ vs $\beta$ when $\rho(0)\to 0$ ($\rho(0)$). We can see three types of phases with increasing $\beta$: first order, hybrid and continuous phase transitions.}
\label{FIG:sextualContact}
\end{figure}

We show in Fig.~\ref{FIG:sextualContact}a the abrupt jump size $J(R_c)$ vs. $\beta$ in of the real sexual contact network and in SI (Fig. S7a and b) for the other networks. One can see that when $\beta$ is large enough, the jump size $J(R_c)$ at the critical resource value $R_c$ diminishes to 0. This means for large $\beta$ values, the infected population size drops continuously with increases in devoted resource (see Fig. S8 in SI), yet discontinuously when $\beta$ is small. The boundary value $\beta_b$ is the lowest $\beta$ value for which the infected population changes continuously vs resource amount at $R_c$ ($J(R_c)$ = 0 in Fig.~\ref{FIG:sextualContact}a and in Fig. S7c for other networks). In SI section II we theoretically analyze the origin of the first order phase transition phenomenon via demonstration in random regular networks.

%\subsection{Non-triviality of Resource Amount and Multi-Phase Behaviours}

\begin{figure}[H]
\centering
 \includegraphics[width=1\linewidth]{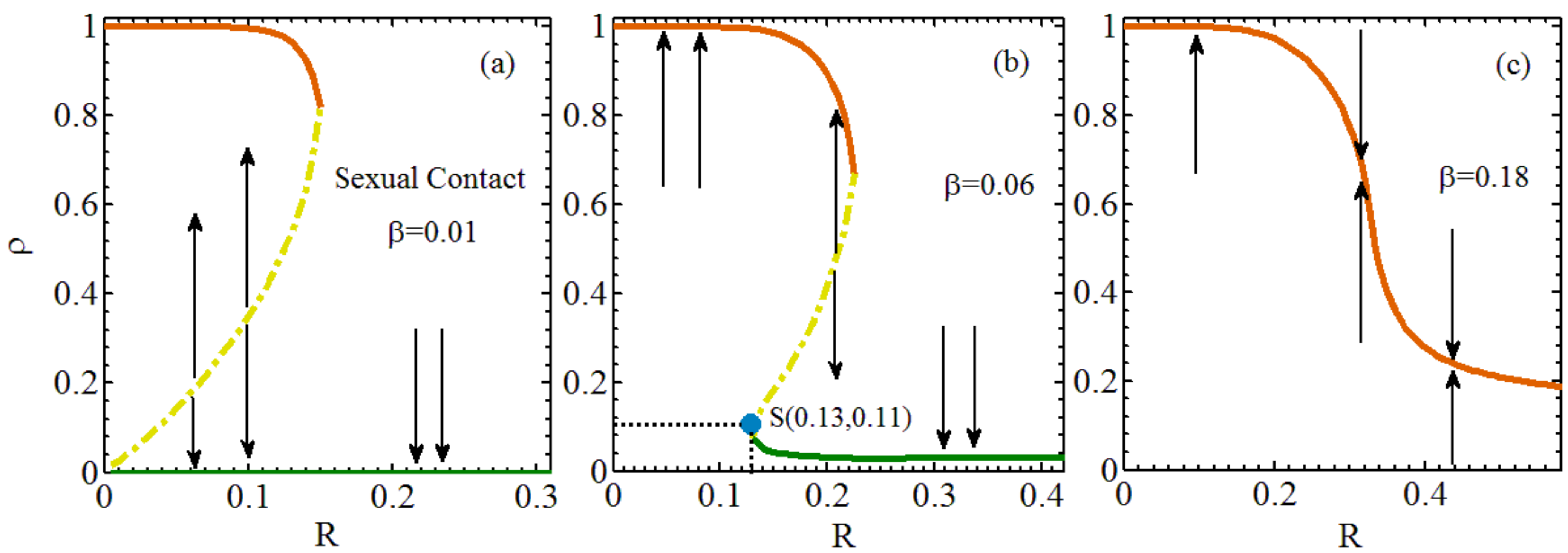}
 \caption{Dynamical stability origin of multi-phase behaviors on the sexual contact network. The curves in the figures are the numerical solutions of $\rho$ (i.e. the average of simulated $p_i(\infty)$ values) in the dynamical Eq.~(\ref{EN:infection}) for different values of $R$. For both (a) the first order transition phase and (b) the hybrid transition phase, the upper orange curve represents the higher steady state solutions, $\rho_h(\infty)$ and the bottom green curve represents the lower solutions, $\rho_l(\infty)$. The middle dashed yellow curve is the unstable steady state solution $\rho_c$ (see Fig. S5 in SI for detailed computation). Therefore if the initial fraction of infected population $\rho(0)$ is above the yellow curve, the system would flow to the higher steady value on the orange curve; Otherwise, the system would flow to the lower steady state value on the green curve. Thus, for each $\rho$ value (initial value $\rho(0)$ to be more accurate) in the plot, the yellow curve defines the critical resource $R_c$. In (b), point $S$ is the lowest point of the yellow dashed line, which means that it leads to a non-zero value of $R_c=0.13$ when initial condition is $\rho(0)\in (0,0.11)$. Such non-zero $R_c$ when $\rho(0) \to 0$ can be observed in Fig.~\ref{FIG:sextualContact}c point $A$ (red). This also explains the observation in Fig.~\ref{FIG:sextualContact}a that the $R_c$ value is flat between $\rho(0)\in(0,0.11)$. For the continuous phase (c), there is only one stable solution of $\rho$. Hence there is no abrupt change in the dynamics.}
\label{FIG:JumpDynamics}
\end{figure}

{\it Our second main finding is regarding heterogenous networks like many real contact networks~\cite{liljeros2001web,liljeros2003sexual,Holme2011Plossimulated} having scale-free degree distributions. In such networks, even if there is a single initial infected node ($\rho(0)\rightarrow0$), a significant amount of resource is required to contain the disease.} This can be seen from Fig.~\ref{FIG:sextualContact}b (and Fig. S9a for other networks), where the variation of $R_c$ v.s. initial infection fraction $\rho(0)$ for the real social network. For many real networks, the value of $R_c$ at $\rho(0)\rightarrow0$ is significantly larger than 0.
%This means even the disease has as few as 1 initial infected node, a considerable amount of resource needs to be set aside to contain it.
This important phenomenon can be understood from Fig.~\ref{FIG:JumpDynamics}b, as the non-zero solution is obtained from the intersection of the stable (green) and unstable solution (yellow) lines (details in caption).
% is caused by that the intersection of the lower stable solutions and the unstable solutions are not 0,  which can be well understood by analysing the dynamical properties of the phase diagram as shown in Fig.~\ref{FIG:JumpDynamics}.

{\it Our third main result is that we find 3 types of phase transitions: first order, hybrid and continuous.} The catastrophic transitions such as the first order and hybrid phase transitions, from spreading dynamics to non-spreading in Eq. (\ref{EN:infection}), are induced by the small spreading probability $\beta$ which usually corresponds to fatal diseases \cite{chowell2014transmission,garnett1998basic}. Fig.~\ref{FIG:sextualContact}c show the 3 distinct phase transitions regimes determined by the value of $\beta$. The three phase regimes are separated by $\beta_c$ and $\beta_b$. Here $\beta_c$ is the critical transmission probability in the original SIS dynamics \cite{pastor2001epidemic} with the recovery rate equal to 1 (corresponding to $R=+\infty$ in our model). In such a scenario, a disease needs to have a $\beta$ larger than $\beta_c$ to spread out. Therefore when $\beta<\beta_c$, increasing resource $R$ can always eradicate  the disease spreading. Hence  $\beta<\beta_c$ is the region of first order phase transition, corresponding to Fig.~\ref{FIG:JumpDynamics}a. When $\beta>\beta_b$, the infected fraction changes continuously with $R$, and we call this region as the continuous phase represented in Fig.~\ref{FIG:JumpDynamics}c. The most interesting region is $\beta_c<\beta<\beta_b$. In this region, since $\beta>\beta_c$, the disease can never be totally removed (even for $R=+\infty$) from the population. At the same time, $\beta<\beta_b$ means as we increase resource $R$, the steady state $\rho(\infty)$ will at some point jump from the upper solution to the lower solution (as shown in Fig.~\ref{FIG:JumpDynamics}b). We refer this region as {\it hybrid phase} (see Fig. S11 for the algorithm details and Fig. S12 for simulation on scale free network).

%\subsection{Phase Transition Regimes}

\begin{figure}[H]
\centering
 \includegraphics[width=0.8\linewidth]{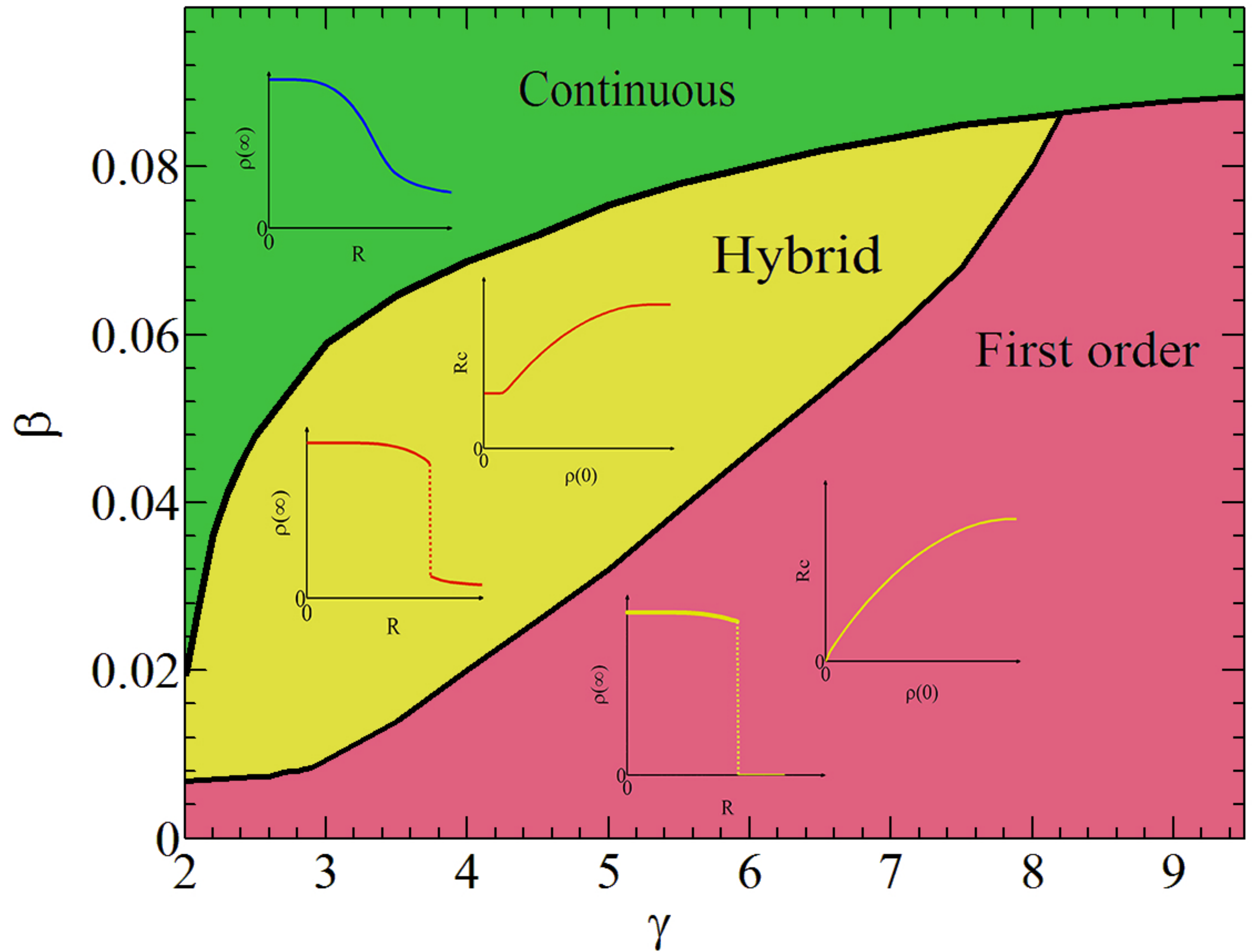}
 \caption{Phase regimes of the system with different infection rate $\beta$ and power law exponent $\gamma$ in scale free networks. Both hybrid and first order phases have abrupt jump in $\rho(\infty)$ when the amount of resource $R$ decreases. The jump in the first order phase starts from 0 whereas the one in the hybrid phase starts from a positive value. As the infection rate increases, the system switches to a continuous phase in which no abrupt jump is observed.}
\label{FIG:Phasediagram}
\end{figure}

Fig.~\ref{FIG:sextualContact}c illustrates the separation of the three distinct phase regimes in sexual contact network (see Fig. S8 and S9 in SI for more information). To investigate the relationship between the presence of these three phases and degree heterogeneity of the network, we carry out further analysis on scale free networks with different  degree distribution exponent $\gamma$, which determines the degree heterogeneity. As shown in Fig.~\ref{FIG:Phasediagram}, for the same value of $\beta$, as the degree heterogeneity decreases (with increasing $\gamma$), the systems evolves from first order to hybrid phase, and eventually to a continuous phase. We notice that when $\gamma$ is small, the three phases (continuous, hybrid and first order) can exist over different values of $\beta$, whereas the hybrid phase disappears for large $\gamma$ (Fig.~\ref{FIG:Phasediagram}) and see Fig. S10 in SI for the results in ER network.

\section{Discussion}

Our results have critical implication on epidemic disease containment, and it is of two folds. First and foremost, the amount of public resource spent in controlling the disease needs to be more than a critical value, otherwise any amount devoted below this level would be wasted without substantial impact on the spreading containment. Additionally, any additional amount of resource far above the critical level would only bring marginal benefit to the containment, as our results indicate that the reduction in final infected population is indeed small. Secondly, because many contact networks such as sexual contact networks and social networks are heterogeneous in their degrees, to effectively contain an epidemic disease, even if the initial number of spreaders is very small, we need to set aside a significant amount of resource. Any hesitation of devoting enough resource will result in cost the lives of many as well as tremendous public resource in the end.

The most fatal diseases measured by the {\it Case Fatality Rate} like Ebola \cite{chowell2014transmission} and HIV \cite{garnett1998basic} are among the least infectious in terms of the {\it Basic Reproduction Number} ($\beta\langle k \rangle$), otherwise they will kill all of their hosts and be unable to reproduce themselves. Therefore, these most fatal diseases tend to have abrupt transitions according to our results, meaning that even a little inadequacy in the devoted public resource could lead to catastrophic outbreaks having considerable casualties. Moreover, during the onset of a disease outbreak, public awareness and disease prevention measures could significantly reduce human contact, which is equivalent to reducing the disease transmission. This will also bring the system to a first order  transition phase. Therefore abrupt changes may be more common and relevant in practice. In such cases it is of extreme importance to investigate the adequacy of resource for fighting the epidemic disease, preventing it from evolving into a large scale pandemic.

%\acknowledgments
%This work was supported by the National Natural Science Foundation of China under Grants Nos. 11575041 and Project No. 61673086.

\bibliography{bibfile,addbibfile}

\end{document}